# Quantum Einstein's Brownian motion

Roumen Tsekov
Department of Physical Chemistry, University of Sofia, 1164 Sofia, Bulgaria

Einstein's Brownian motion of a quantum particle in a classical environment is studied via virial and equipartition theorems. The effect of continuous measurement in a strongly dissipative environment is accounted for and a quantum generalization of the classical Einstein law of Brownian motion is obtained. A thermo-quantum Smoluchowski diffusion equation is derived via a generalization of the Madelung quantum hydrodynamics. The latter is applied for description of the quantum tunneling at equilibrium and stationary states as well as of the motion of an electron in metals, i.e. the Smoluchowski-Poisson problem.

In 1905 Einstein has published three papers, which revolutionized our knowledge for the Nature. In the first one, accepting that the light consists of discrete quanta of radiation, called later photons, he explained the photoelectric effect that certain metals emit electrons when irradiated. For this contribution to quantum theory Einstein received the 1921 Nobel Prize in Physics. In the second paper Einstein proposed the special theory of relativity, which reinterprets the classical mechanics presuming that the speed of light remains constant in all frames of reference. Thus he discovered the equivalence of mass and energy. The third of his seminal papers [1] concerned the Brownian motion, where Einstein calculated the average trajectory of a microscopic particle forced by random collisions with the molecules in a fluid. He extended the Boltzmann view and provided convincing evidence for the physical existence of molecules. His probabilistic theory let to enormous progress in our understanding of non-equilibrium thermodynamics and is a precursor of many modern kinetic models of diffusive stochastic processes in the Nature.

According to the classical statistical mechanics the equilibrium in the momentum space is achieved when a particle spent in a fluid a period larger than the classical momentum relaxation time $m/b$, where $m$ and $b$ are the particle mass and friction coefficient, respectively. Thus, following the equipartition law the particle momentum dispersion $\sigma_p^2 = mk_BT$ is proportional to the temperature $T$. On the other hand, the virial theorem states that at large times the momentum dispersion can be expressed by the force acting on the particle via the relation $\sigma_p^2/m = -<Fx>$. After Langevin [2] the total force $F = -b\dot{x} + f$ is a sum of the friction force, proportional to the Brownian particle velocity, and the stochastic Langevin force $f$. Substituting this expression the virial theorem acquires the form $\sigma_p^2 = b\sigma_{xp}$, since the random contribution of $f$ vanishes because the Langevin force is not correlated to the position of the Brownian particle. Integrating now on time the virial theorem accomplished by the Maxwell expression $\sigma_p^2 = mk_BT$ yields the Einstein law of Brownian motion

$$\sigma_x^2 = 2Dt \tag{1}$$

where the particle diffusion constant $D = k_BT/b$ is introduced via the classical Einstein formula. This linear dependence of the particle position dispersion $\sigma_x^2$ on time $t$ is proven many times and has become a textbook law for diffusion processes.

After the establishment of the Einstein theory of Brownian motion the quantum mechanics was born. Although it changed drastically the description of classical mechanics, the Einstein law (1) is still regularly applied to quantum objects. Indeed, the fluctuation-dissipation theorem [3] shows that the fluctuating Langevin force in a quantum environment is not a white noise but many of the important examples in practice are for a quantum Brownian particle moving in a classical environment. For instance, in the case of an electron moving in a crystal the lattice vibrations can be described via the classical mechanics down to relatively low temperatures. Colored noises are observed also in classical systems and naturally reflect in deviations from the Einstein law but they are out of the scope of the present paper. The quintessence of quantum mechanics is the Heisenberg uncertainty principle [4], which imposes a restriction on the momentum and position dispersions of a quantum particle. Substituting in the Robertson-Schrödinger uncertainty relation $\sigma_x^2\sigma_p^2 - \sigma_{xp}^2 \geq \hbar^2/4$ [5, 6] the classical Einstein expressions $\sigma_p^2 = mk_BT$, $\sigma_{xp} = mD$ and $\sigma_x^2 = 2Dt$ yields immediately that the Einstein law (1) satisfies the Heisenberg principle only at large times, $t \geq \lambda_T^2/2D + m/2b$ where $\lambda_T \equiv \hbar/2\sqrt{mk_BT}$ is the thermal de Broglie wave length. Since the Einstein law is valid for large times $t \geq m/b$, a necessary condition for consistency of these inequalities is $\lambda_T^2/D \leq m/b$. The latter shows that the uncertainty principle holds for a quantum Einstein's Brownian particle at faster classical diffusion only, $D \geq \hbar/2m$. The quantum characteristic time $\lambda_T^2/D$ can be also written as an oscillator relaxation time $b/m\omega_2^2$ with the second Matsubara frequency $\omega_2 = 2k_BT/\hbar$.

Obviously, the Einstein's Brownian motion of a quantum particle in a classical environment is not described by Eq. (1) and the scope of the present analysis is to obtain a generalization of the classical Einstein law. This is especially important at low temperature and strong friction, where $D < \hbar/2m$. If a Gaussian wave packet spreads in vacuum the constant momentum dispersion $\sigma_p^2 = \hbar^2/4\sigma_x^2(0)$ is determined by the initial position dispersion of the packet. These momentum and position dispersions are due to an initial measurement fixing the wave packet. In highly dissipative environment the quantum particle is continuously measured by the environment and, for this reason, any moment could be considered as a new beginning. Hence, the momentum dispersion $\sigma_p^2 = \hbar^2/4\sigma_x^2$ in this case will depend on the current value of the position dispersion. This is the minimal Heisenberg relation, which is valid for Gaussian processes as free Brownian motion certainly is. Substituting this expression into the virial theorem $\sigma_p^2 = b\sigma_{xp}$ being valid for quantum objects as well, and integrating on time yields an expression for the purely quantum diffusion [7]

$$\sigma_x^2 = \hbar\sqrt{t/mb} \tag{2}$$

Since Eq. (2) is valid for $t > m/b$ it follows that $\sigma_x^2 > \hbar/b$ and $\sigma_p^2 = \hbar\sqrt{mb/t}/4 < \hbar b/4$. Considering now the case of a non-zero temperature, the thermalization of the Brownian particle is achieved at $t > m/b$ indeed but quantum relaxations are still in progress since $\sigma_x^2$ is finite. Hence, in this case one can propose the following Maxwell-Heisenberg expression for the particle momentum dispersion

$$\sigma_p^2 = mk_BT + \hbar^2/4\sigma_x^2 \tag{3}$$

which is a superposition between the thermal and quantum components. It is valid at high friction only, satisfies the Heisenberg principle at any time and reduces at infinite time to the equilibrium results $\sigma_p^2 = mk_BT$ and $\sigma_x^2 = \infty$. Obviously due to quantum effects the momentum relaxation of a quantum Brownian particle is non-exponential. The thermo-quantum expression (3) corresponds to continuous measurements of the quantum Brownian particle by a thermal dissipative environment. Introducing Eq. (3) into the virial theorem $\sigma_p^2 = b\sigma_{xp}$ results after integration on time in the following generalization of the Einstein law of Brownian motion [7]

$$\sigma_x^2 - \lambda_T^2 \ln(1 + \sigma_x^2/\lambda_T^2) = 2Dt \tag{4}$$

At large time $\sigma_x^2$ is larger than the thermal de Broglie wave length square $\lambda_T^2$ and hence Eq. (4) tends asymptotically to Eq. (1). At short time, however, Eq. (4) reduces to Eq. (2), which describes the purely quantum diffusion at zero temperature as well.

It is interesting to explore how the classical diffusion equation will change for a quantum Brownian particle. Traditionally, the quantum particles are described by the Schrödinger equation. In 1927 Madelung has demonstrated [8], however, that the Schrödinger equation can be presented in an alternative hydrodynamic form

$$\partial_t \rho = -\nabla \cdot (\rho V) \qquad\qquad m\partial_t V + mV \cdot \nabla V = -\nabla U - \nabla \cdot \mathbb{P}_Q / \rho \qquad (5)$$

where $\rho$ is the probability density, $V$ is the hydrodynamic-like velocity, $U$ is an external potential and $\mathbb{P}_Q \equiv -(\hbar^2/4m)\rho \nabla \otimes \nabla \ln \rho$ is a quantum pressure tensor. These Euler equations describe the evolution of the probability density in vacuum. In the case of a classical dissipative environment one can enhance the dynamic balance (5) by a friction force $-bV$ and thermal pressure tensor $k_B T \rho \mathbb{I}$ to obtain the equation $m\partial_t V + mV \cdot \nabla V + bV = -\nabla(U + Q + k_B T \ln \rho)$ [7], where $Q \equiv -\hbar^2 \nabla^2 \sqrt{\rho}/2m\sqrt{\rho}$ is the Bohm quantum potential [9], being related to the quantum pressure tensor via the Gibbs-Duhem relation $\nabla \cdot \mathbb{P}_Q = \rho \nabla Q$. Neglecting now the inertial terms at strong friction this results in

$$V = -\nabla(U + Q + k_B T \ln \rho)/b \qquad (6)$$

Introducing the velocity $V$ from Eq. (6) into the first continuity equation from Eqs. (5) yields the following thermo-quantum diffusion equation [7]

$$\partial_t \rho = \nabla \cdot [\rho \nabla (U + Q)/b + D\nabla \rho] \qquad (7)$$

This is a Smoluchowski equation describing classical diffusion in the fields of an external and quantum potentials. The latter is, however, a non-trivial potential and depends on the probability density. It is, in fact, a chemical potential originating from the particle quantum kinetic energy. While the logarithmic term in Eq. (6) originates from the Boltzmann entropy, the average value of the quantum potential is proportional to the Fisher entropy. The solution of Eq. (7) for a free particle is a Gaussian probability density with position dispersion given by Eq. (4).

One of the most interesting quantum phenomena is the tunneling effect. If an equilibrium system is considered than $V = 0$ and Eq. (6) reduces after integration to the following differential equation regarding the function $\ln \rho_{eq}$

$$-\hbar^2 (\nabla^2 \ln \rho_{eq})/4m - \hbar^2 (\nabla \ln \rho_{eq})^2 /8m + k_B T \ln \rho_{eq} = F - U \qquad (8)$$

Here the quantum potential is presented in an alternative form and $F$ is the constant free energy. This equation provides several important results. In the classical limit, for instance, the probability density is the Boltzmann distribution $\rho_{eq} = \exp[(F - U)/k_B T]$. At zero temperature and a strong potential $(U > E)$ the second quadratic term in Eq. (8) dominates and thus Eq. (8) provides the well-known tunneling WKB distribution

$$\rho_{eq} = C\exp[-2\int dx\sqrt{2m(U-E)}/\hbar]$$

where $E$ is the full energy. An interesting new result follows from Eq. (8) for a weak potential ($U < E$) at zero temperature, when the leading term is the first one

$$\rho_{eq} = C\exp[4m\int dx\int dx(U-E)/\hbar^2]$$

Therefore, in case of a general potential the solution of the complete Eq. (8) is required.

    The dynamics of the tunneling effect can be also described by the Madelung hydrodynamics. Neglecting the non-stationary term in the dynamic Eq. (5) and integrating the resultant equation yields the stationary hydrodynamic velocity $V_x = \sqrt{2(E-U-Q)/m}$. Now introducing it in the first continuity equation from Eq. (5) leads to the following equation

$$\partial_t \rho = -\partial_x[\rho\sqrt{2(E-U-Q)/m}] \qquad (9)$$

describing the dynamics of quantum tunneling as well. The stationary solution of Eq. (9) is given by the differential equation

$$\rho_{st} = C/\sqrt{2(E-U)/m + \hbar^2 \partial_x^2 \sqrt{\rho_{st}}/m^2\sqrt{\rho_{st}}}$$

which in the classical limit reduces to the ergodic distribution $\rho_{st} = C/\sqrt{2(E-U)/m}$, following from the relative time spent at a given position. Substituting of this classical distribution in the quantum potential above results in a semiclassical stationary distribution

$$\rho_{st} = C/\sqrt{2(E-U)/m + \hbar^2 \partial_x^2 U/4m^2(E-U) + 5\hbar^2(\partial_x U)^2/16m^2(E-U)^2}$$

Naturally, in the equilibrium case Eq. (9) reduces to Eq. (8) at zero temperature. The effect of temperature and friction on the quantum tunneling can be described by the dissipative Madelung hydrodynamics [7] discussed before. Moreover, the Madelung hydrodynamic approach provides possibility for studying nonlinear friction [10] and relativistic [11] effects on the quantum tunneling as well.

    The quantum Smoluchowski equation (7) can be employed also for description of the motion of an electron in a metal under the action of an electric potential $\phi$. Since at large friction the electron velocity is small the electric potential can be described well via the electrostatic Poisson equation

$$\varepsilon_0\varepsilon\nabla^2\phi = e(\rho-\rho_0) \qquad (10)$$

where $\varepsilon_0\varepsilon$ is dielectric permittivity and the density $\rho_0$ of the positive charge in the metal is accepted to be constant in the frames of the jelly model. If the electron density $\rho$ differs slightly from the equilibrium value one can linearize Eq. (7) with $U = -e\phi$ around $\rho_0$ to obtain

$$\partial_t\rho = \nabla\cdot(-e\rho_0\nabla\phi/b - \hbar^2\nabla^3\rho/4mb + D\nabla\rho) \qquad (11)$$

Excluding now the electron density $\rho$ among the system of linearized quantum Smoluchowski-Poisson equations (10) and (11) leads to a diffusive equation for the electric potential

$$\partial_t\phi = -D(\phi/\lambda_D^2 + \lambda_T^2\nabla^4\phi - \nabla^2\phi) \qquad (12)$$

where $\lambda_D \equiv \sqrt{\varepsilon_0\varepsilon k_B T/e^2\rho_0}$ is the Debye screening length. Using Eq. (10), Eq. (12) can be rewritten for the electron density as well

$$\partial_t\rho = -D[(\rho-\rho_0)/\lambda_D^2 + \lambda_T^2\nabla^4\rho - \nabla^2\rho] \qquad (13)$$

The Fourier image of the solution of Eq. (13) reads

$$\rho_q = \rho_0[1-\exp(-Dt/\lambda_D^2)]\delta(q) + \exp(-D_q q^2 t) \qquad (14)$$

where the effective diffusion coefficient equals to $D_q = D(1/\lambda_D^2 q^2 + 1 + \lambda_T^2 q^2)$. For small wave vectors the diffusion is driven mainly by the electrostatics and $D_q \approx D/\lambda_D^2 q^2 = e^2\rho_0/\varepsilon_0\varepsilon b q^2$. At large wave vectors the electron is very localized and the diffusion is dominated by quantum effects with $D_q \approx D\lambda_T^2 q^2 = \hbar^2 q^2/4mb$. In the intermediate region of wave vectors the effective diffusion coefficient exhibits a minimum $D_q = D(1+2\lambda_T/\lambda_D) = (k_B T + \hbar\omega_0)/b$ at $q^2 = 1/\lambda_D\lambda_T$, where $\omega_0 = \sqrt{e^2\rho_0/m\varepsilon_0\varepsilon}$ is the Langmuir plasma frequency. Since the Smoluchowski-Poisson problem describes also other systems [12], the present analysis is relevant to the significance of quantum effects there, for instance, in the self-gravitation.

In the case of Earth gravitation with the constant acceleration $g$ Eq. (7) reduces to

$$\partial_\tau\rho = \partial_\zeta\{\rho\partial_\zeta[\zeta - 2(\alpha_T\lambda_T)^2\partial_\zeta^2\sqrt{\rho}/\sqrt{\rho}] + \partial_\zeta\rho\} \qquad (15)$$

where the classical dimensionless distance $\zeta \equiv \alpha_T z$ and time $\tau \equiv D\alpha_T^2 t$ with $\alpha_T \equiv mg/k_B T$ are introduced. As is seen the importance of the quantum term increases by the increase of the ratio of the thermal de Broglie length $\lambda_T$ and the classical mean barometric height $1/\alpha_T$. Their equality defines via $\alpha_T \lambda_T = 1$ a characteristic temperature $T_g = \sqrt[3]{2mg^2\hbar^2}/2k_B$, which is quantum but increases by the particle mass $m$. For electrons, for instance, it equals to $T_g = 0.45$ nK, which corresponds to a macroscopic thermal de Broglie length $\lambda_{T_g} = 0.7$ mm. One can present generally the solution of Eq. (15) as a sum of the classical barometric distribution and a quantum correction, $\rho = \alpha_T \exp(-\zeta) + w$. If the latter is small one can linearize Eq. (15) to obtain

$$\partial_\tau w = \partial_\zeta w - (T_g/T)^3 (2\partial_\zeta^4 w + 3\partial_\zeta^3 w + \partial_\zeta^2 w)/2 + \partial_\zeta^2 w \tag{16}$$

This equation describes the combined effect of biharmonic, cubic and harmonic quantum diffusions to the classical Smoluchowski equation, which is strongly depressed by the temperature.

Finally, let us draw a simple picture of the quantum Einstein's Brownian motion, which acts in two steps: a quantum and a classical one. First, it performs a standard 3D Brownian walk with a universal quantum diffusion coefficient $\hbar/2m$ [13]. Hence, following Einstein we have for the particle dispersion the expression

$$\sigma_x^2 = \hbar \tau / m \tag{17}$$

where $\tau$ is the duration of this movement. The length of the path drawn by the particle equals to $l = c\tau$, where $c$ is the real velocity of the particle, e.g. $c^2 = k_B T/m$. Thus, Eq. (17) acquires the form, which is also known from the polymer physics,

$$\sigma_x^2 = \hbar l / mc \tag{18}$$

Hence, the segment length equals to the de Broglie one $\hbar/mc$. Secondly, we put the particle to perform 1D classical Brownian movement along the fractal trajectory drawn by the first quantum move. Similar problem is already discussed in the literature [14]. According to Einstein the particle will reach the path end at time $t = l^2/D$, where $D$ is the usual classical Einstein diffusion coefficient, e.g. $D = k_B T/b = mc^2/b$. Substituting $l$ from this relation in Eq. (18) yields

$$\sigma_x^2 = \hbar\sqrt{t/mb} \tag{19}$$

This expression coincides with Eq. (2), being obviously valid for $\hbar/2m > D$, since otherwise the second step can outrun the first one. If this happens, the classical trajectory will sometimes lead the simultaneous quantum one: in this case the thermo-quantum diffusion will appear which is described by Eq. (4). Note that at $t \to \infty$ the Einstein law holds, which is due to junctions in the quantum trajectory. The present analysis shows that the quantum Brownian particle undergoes a classical Brownian motion in a Brownian-fluctuating quantum space. This idea is somehow close to the de Broglie pilot-wave, which corresponds to the fist quantum move drawing the path, along which the particle to perform the second classical Einstein's Brownian motion.

[1]   A. Einstein, Ann. Phys. (Leipzig) **17**, 549 (1905)
[2]   P. Langevin, Comp. Rend. Acad. Sci. (Paris) **146**, 530 (1908)
[3]   H.B. Callen and T.A. Welton, Phys. Rev. **83**, 34 (1951)
[4]   W. Heisenberg, Z. Phys. **43**, 172 (1927)
[5]   H.R. Robertson, Phys. Rev. **35**, 667A (1930)
[6]   E. Schrödinger, Berg. Kgl. Akad. Wiss. (Berlin) **24**, 296 (1930)
[7]   R. Tsekov, Int. J. Theor. Phys. **48**, 85 (2009), ibid. **48**, 630 (2009)
[8]   E. Madelung, Z. Phys. **40**, 322 (1927)
[9]   D. Bohm, Phys. Rev. **85**, 166 (1952)
[10]  R. Tsekov, Ann. Univ. Sofia, Fac. Phys. **105** (2012) 000 arXiv 1003.0304
[11]  R. Tsekov, Ann. Univ. Sofia, Fac. Phys. **105** (2012) 000 arXiv 1003.2693
[12]  P.H. Chavanis, Phys. Rev. E **68**, 036108 (2003)
[13]  E. Nelson, Phys. Rev. **150**, 1079 (1966)
[14]  B.P. Radoev and B.G. Tenchov, J. Phys. A: Math. Gen. **20**, L159 (1987)
[15]  A.B. Datsev, Quantum Mechanics, Science and Art, Sofia (1963), in Bulgarian
[16]  A.B. Datzeff, J. Phys. Rad. **20**, 949 (1959); **21**, 201 (1960); **22**, 35, 101 (1961); **23**, 241 (1962)
[17]  A.B. Datzeff, Nuovo Cim. B **29**, 105 (1975)
[18]  A.B. Datzeff, Phys. Lett. A **59**, 185 (1976)
[19]  A.B. Datseff, Int. J. Quant. Chem. **28**, 739 (1985)
[20]  A.B. Datzeff, in Open Questions in Quantum Physics, eds. G. Tarozzi and A. van der Merwe, Reidel, Dordrecht (1985), p. 215
[21]  A.B. Datzeff, in Microphysical Reality and Quantum Formalism, eds. A. van der Merwe, F. Selleri and G. Tarozzi, Kluwer, Dordrecht (1988), p. 195